\documentclass{iau}
\usepackage{graphicx}

\title[Synp 289~~The Heliometer of the Observat\'orio Nacional] 
{Development and First Year of Results from the Heliometer of Observat\'orio Nacional}

\author[Andrei \it{et al.}]   
{Alexandre H. Andrei$^1$
  \thanks{CNPq grant PQ306775/2009-3 and PARSEC International Incoming Fellowship of the Marie Curie 7th ECP. Present address: Shanghai Astronomical Observatory, CAS, China.},
 \ Victor A. D'\'Avila$^{2,3}$,
 \ Eug\^enio Reis Neto$^4$,
 \ Jucira L. Penna$^2$,
 \ S\'ergio C. Boscardin$^2$,
 \ Alissandro Coletti$^5$,
 \\\ Luiz C. Oliveira$^6$
 \and Costantino Sigismondi$^7$}

\affiliation{$^1$(1) Observat\'orio Nacional/MCTI-Brasil; (2) Osservatorio Astrofisico di Torino/INAF-Italia; (3) SYRTE/Observatoire de Paris-France; (4) Observat\'orio do Valongo/UFRJ-Brasil \\ email: {\tt oat1@on.br} \\[\affilskip] 
$^2$Universidade Estadual do Rio de Janeiro, Brasil
$^3$Observat\'orio Nacional/MCTI, Brasil
$^4$Museu de Astronomia e Ci\^encias Afins/MCTI, Brasil
$^5$Azeheb/FACC, Brasil
$^6$NGC51, Brasil
$^7$ICRA/Universi\`a Pontificia Regina Apostolorum/Sapienza Universit\`a di Roma, Italia}

\pubyear{2012}
\volume{294}  
\pagerange{119--126}
\jname{Solar and Astrophysical Dynamos and Magnetic Activity}
\editors{A.G. Kosovichev, E.M. de Gouveia Pinto \& Y. Yan, eds.}
\begin{document}

\maketitle

\begin{abstract}
Recent research on global climate changes points to three distinct sources of climate disturbance: anthropogenic; natural changes in the oceans and atmosphere; and irregularities in the solar cycles. One of the most direct ways to survey an exogenous component of the climatic variability is through the measurement of variations in the diameter and shape of the solar disk.
At Observat\'orio Nacional/MCTI, Rio de Janeiro, after several years of diameter observation using a CCD Solar Astrolabe, these measurements are now performed by a state-of-the-art Solar Heliometer. The heliometric method is one of the most successful techniques to measure small variations of angles. Its principle has been used for the latest space borne astrometric missions, aiming to milli-arcsecond precision. The success of this method relies in the fact that it minimizes the dependence of angular measurements to the thermal and mechanical stability of the instrument. However in the classic heliometer the objective is split into two halves to which is applied a linear displacement along the cut, thus still leaving room for a residual dependence with the focus, due to non-concentricity of the beams of the two images. The focus variation, as well as the effects brought by the large temperature variations during solar observations, was tackled in the Solar Heliometer by having all optical elements and their niches made on CCZ, and the telescope tube on carbon fiber, both materials of negligible thermal coefficient. Additionally, the measures are made perpendicular to the separation direction and the plate scale can be known at every time from the solar motion itself. 
We present the results from the first year of measurements, in special exploring the upheaval of solar activity on late 2011.
\keywords{solar diameter, heliometer, cycle 24}
\end{abstract}

\firstsection 
\section{The Development and Realization of the Heliometer}

At Brazil's Observat\'orio Nacional, in Rio de Janeiro, regular measurements of
solar diameter variations started in 1997 with a CCD astrolabe. The series
  extended till 2009 with high density of daily measures ($\sim$20/day) (Penna et al. 2010). 
Aiming to attain precision at 100$\it{mas}$ for a single measure and fully daily
coverage on heliolatitude, the Heliometer was developed.

The heliometric mirror is made of CCZ-HS, a ceramic material with very low thermal
  expansion coefficient (0.0 $\pm$ 0.2 $\times$ 10$^{-7}$/$^\circ$C, in the range 0-50$^\circ$C). 
 The two half mirrors are immobilized, in relation to each other, by means of a external
  ring, all resting over an optical plate. Its cell guarantees the mechanical and
  geometrical stability for the entire set. This niche is also in CCZ. 
 The surface quality of the optical plate and the mirrors is better than $\lambda$¿/12 and $\lambda$/20,
  respectively. 
 A mask at the top of the cell has been designed to keep the two half mirrors blocked in
  place and also to assure that the entrance pupil has a symmetric shape, regularizing
  the PSF. 
 The tube of the telescope is made of carbon fiber. This material, as well as extremely
  rigid, has very low coefficient of thermal expansion. It is mounted inside a stainless
  steel truss support and can rotate around its axis. 
 In order to eliminate the secondary mirror the CCD chip was removed apart from the camera electronics and installed directly in the focal plane. 
 Each half-mirror is tilted of an angle slightly greater than 0.135$^\circ$ in order to displace the images relatively to each other by one solar diameter approximately. In this way we will have the opposite limbs of the Sun almost in tangency in the focal plane at the perihelion. 

\section{First Results}

The field tests included: the following of the 2010 total Solar eclipse, in which the Heliometer was used as such in between the contacts while at those instants performing in the classical way; a nearly 2 months campaign of zenithal measurements, thus free of refraction terms, as enabled by the -22.9$^\circ$ latitude of Rio de Janeiro; and simultaneous measurements, at the same wavelength window, of the Venus transit in 2012. All the experiments confirmed the instrument intrinsic precision at 100$\it{mas}$.

The first year series reveal a mounting diameter. This is compatible with the earlier CCD astrolabes rprediction of an increase that combines a variation tied on-phase to the 11y-cycle, which reflects the late 2011 upheaval of cycle 24, plus a variation tied antiphase to a Gleissberg-like cycle, which reflects the decreasing amplitude of the last 2 cycles. On the short term, the Heliometer series already shows a good time coincidence of local diameter maxima with the major CME events on 2011.


\begin{figure}[ht!]
\begin{minipage}[b]{0.45\linewidth}
\begin{center}
 \includegraphics[width=\textwidth]{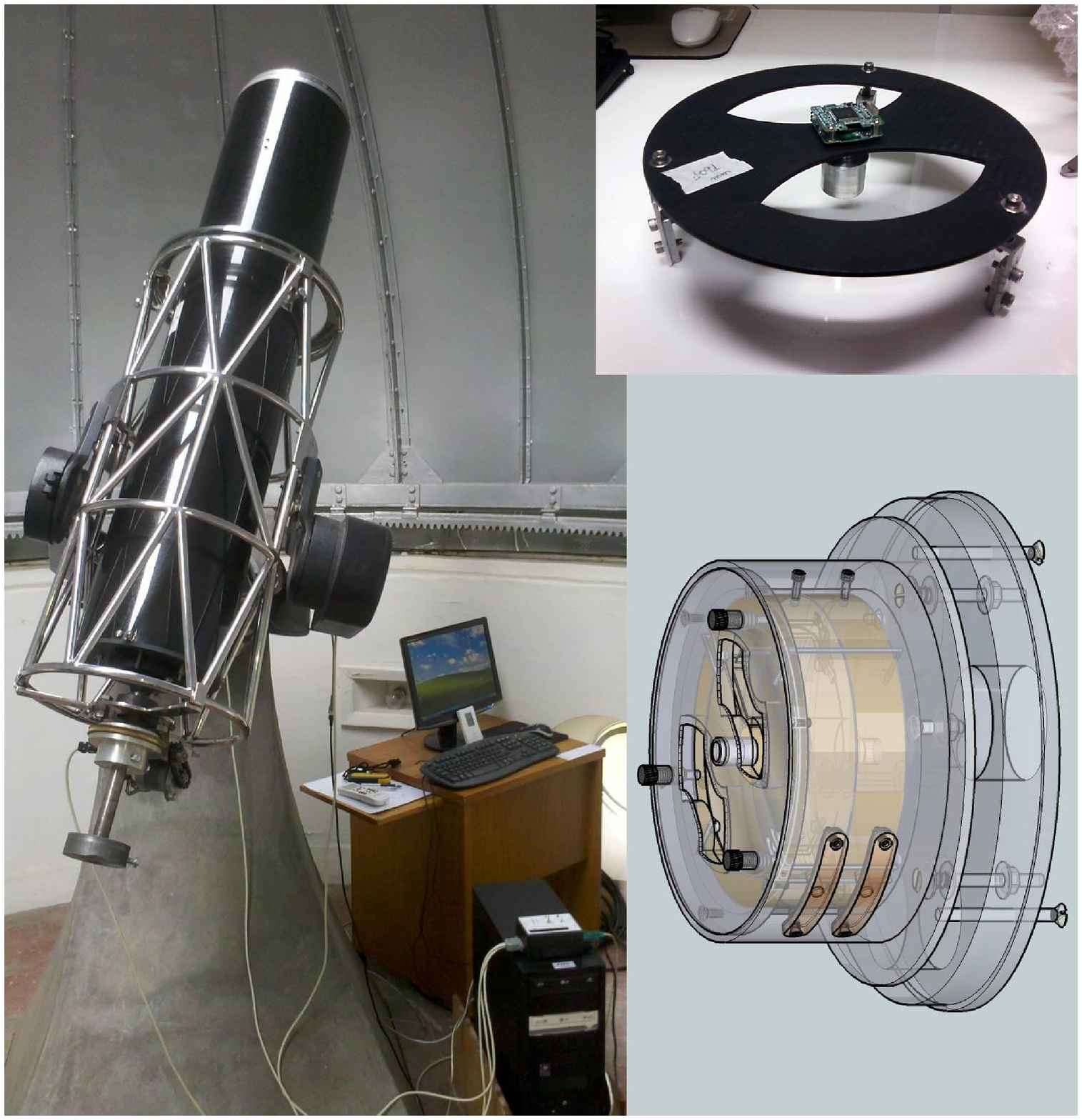}
   \label{fig1}
\end{center}
\end{minipage}
\begin{minipage}[b]{0.55\linewidth}
\begin{center}
 \includegraphics[width=\textwidth]{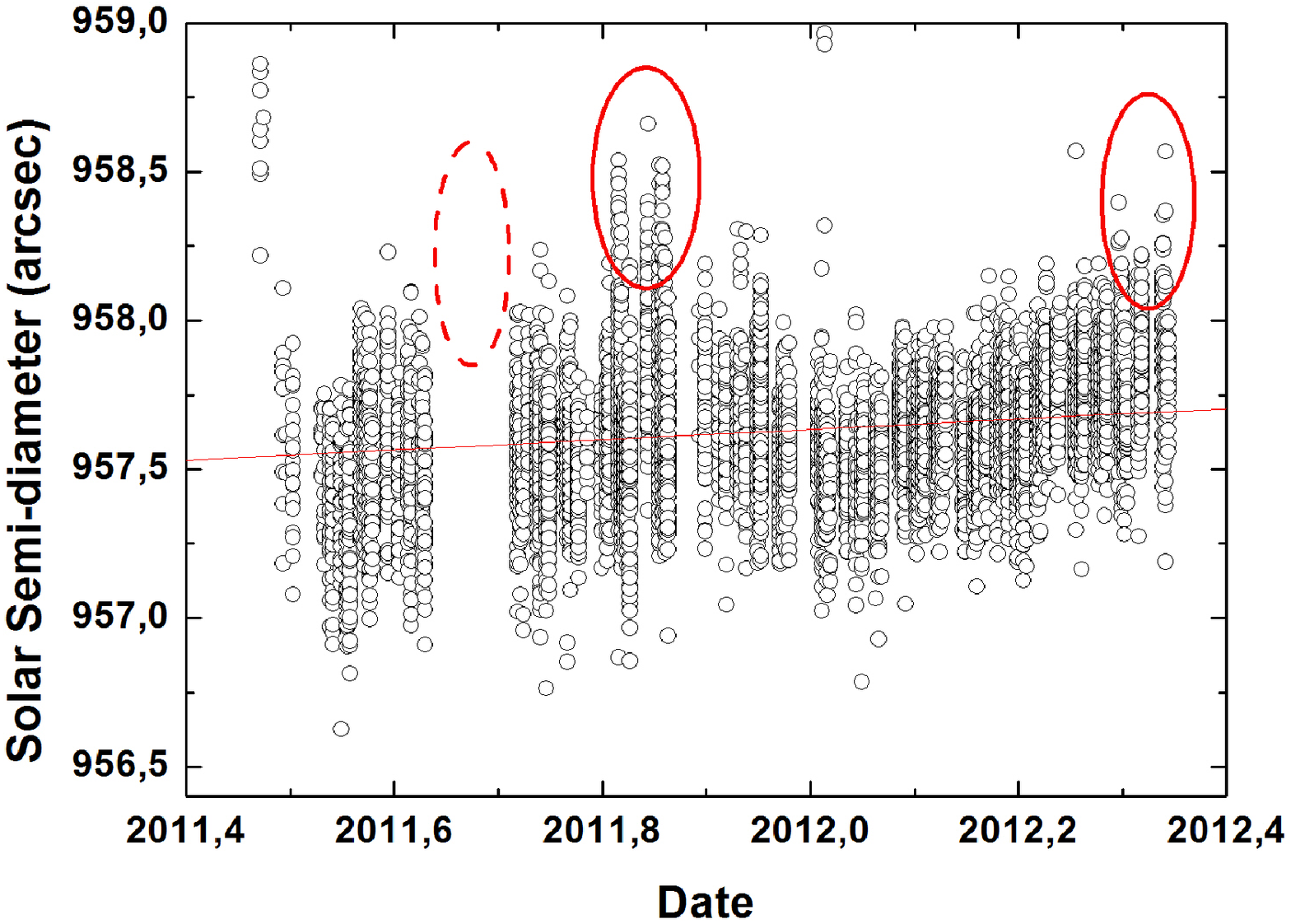}
 \caption{At left, the Heliometer, plus views of the diffraction mask with the CCD and of the hemi-mirrors and niche. On top, the Heliometer first year raw series; the red regions are those when the largest groups of CMEs happened in 2011, which coincides with local maxima of the measured semi-diameter.} 
   \label{fig2}
\end{center}
\end{minipage}
\end{figure}






\end{document}